\documentclass[sigconf]{acmart}
\AtBeginDocument{%
  }

\setcopyright{acmlicensed}
\copyrightyear{2026}
\acmYear{2026}
\acmDOI{XXXXXXX.XXXXXXX}
\acmConference[KDD '26]{The 32nd ACM SIGKDD Conference on Knowledge Discovery and Data Mining}{August 9--13, 2026}{Jeju, Korea}

\usepackage{booktabs}
\usepackage{graphicx}

\begin{document}

\title{Decision Quality Evaluation Framework at Pinterest}


\author{Yuqi Tian}
\affiliation{%
  \institution{Pinterest}
  \city{San Francisco, CA}
  \country{USA}}
\email{yuqitian@pinterest.com}

\author{Robert Paine}
\authornote{Work done at Pinterest.}
\affiliation{%
  \institution{Pinterest}
  \city{San Francisco, CA}
  \country{USA}}
\email{rpaine@pinterest.com}

\author{Attila Dobi}
\affiliation{%
  \institution{Pinterest}
  \city{San Francisco, CA}
  \country{USA}}
\email{adobi@pinterest.com}

\author{Kevin O'Sullivan}
\affiliation{%
  \institution{Pinterest}
  \city{New York, NY}
  \country{USA}}
\email{kosullivan@pinterest.com}

\author{Aravindh Manickavasagam}
\affiliation{%
  \institution{Pinterest}
  \city{Richmond, VA}
  \country{USA}}
\email{amanickavasagam@pinterest.com}

\author{Faisal Farooq}
\affiliation{%
  \institution{Pinterest}
  \city{Murphy, TX}
  \country{USA}}
\email{ffarooq@pinterest.com}

\renewcommand{\shortauthors}{Tian et al.}

\begin{abstract}
  Online platforms require robust systems to enforce content safety policies at scale. A critical component of these systems is the ability to evaluate the quality of moderation decisions made by both human agents and Large Language Models (LLMs). However, this evaluation is challenging due to the inherent trade-offs between cost, scale, and trustworthiness, along with the complexity of evolving policies. To address this, we present a comprehensive Decision Quality Evaluation Framework developed and deployed at Pinterest. The framework is centered on a high-trust Golden Set (GDS) curated by subject matter experts (SMEs), which serves as a ground truth benchmark. We introduce an automated intelligent sampling pipeline that uses propensity scores to efficiently expand dataset coverage. We demonstrate the framework's practical application in several key areas: benchmarking the cost-performance trade-offs of various LLM agents, establishing a rigorous methodology for data-driven prompt optimization, managing complex policy evolution, and ensuring the integrity of policy content prevalence metrics via continuous validation. The framework enables a shift from subjective assessments to a data-driven and quantitative practice for managing content safety systems.
\end{abstract}



\received{20 February 2007}
\received[revised]{12 March 2009}
\received[accepted]{5 June 2009}

\maketitle

\section{Introduction}

User trust is paramount to the success of online platforms. At Pinterest, our goal is to cultivate a safe and inspiring environment for our users. To achieve this, we establish comprehensive content safety policies and community guidelines across various policy areas \cite{pinterest_guidelines_2025}. These broad principles are then translated into detailed internal enforcement policies for our operational teams. This complex policy landscape requires scalable enforcement through a combination of machine learning models, human agent labeling, and, increasingly, Large Language Models (LLMs). The integrity of decisions guided by these internal policies directly impacts user safety and trust.

The broader challenge of platform governance at scale is a significant area of research, encompassing the technical and political complexities of algorithmic content moderation \cite{gorwa2020algorithmic}. Studies have explored the critical dynamics of human-machine collaboration in these moderation systems \cite{jhaver2019human} and conducted large-scale empirical investigations into how community norms are enforced on major platforms \cite{chandrasekharan2018internet}. While this prior work establishes the operational and social challenges of moderation, our paper addresses a complementary and foundational need: the development of a rigorous and scalable evaluation framework to measure and ensure the quality and consistency of those moderation decisions.

\begin{figure}
    \centering
    \includegraphics[width=\linewidth]{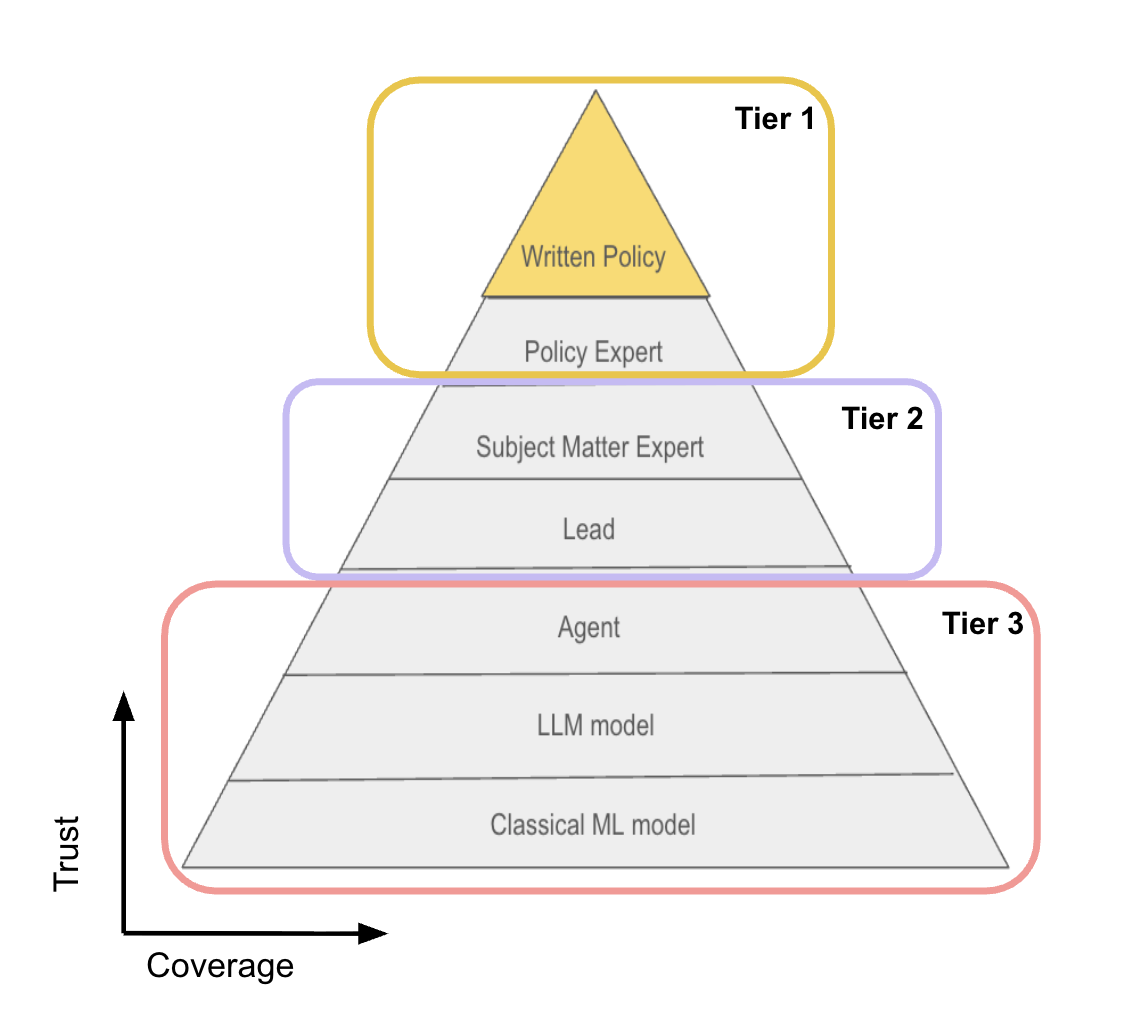}
    \caption{The Hierarchy of Trustworthiness for Labeling Sources, illustrating the trade-off between label quality and scalability. \textbf{Policy Experts} are the authors and ultimate interpreters of the written policy. \textbf{SMEs} are highly-trained specialists who produce the highest-quality labels for creating ground truth datasets like the GDS. \textbf{Leads} are experienced reviewers who oversee the quality of the scalable agent workforce at the base.}
    \label{fig:pyramid}
\end{figure}

The use of Generative AI has become critical in our content safety operations, from prevalence measurement \cite{pinterest_prevalence_2025} to complex policy interpretation. These powerful systems, however, must operate within a landscape defined by several related challenges. A significant challenge in the Content-Safety domain is the inherent ambiguity and complexity of policies, which leads to inconsistent application by both human and automated agents. This is compounded by the extreme rarity of content and the high cost associated with obtaining high-quality labels from subject matter experts (SMEs). These pressures give rise to a fundamental trade-off between trustworthiness, scale, and cost, a concept we term the ``Pyramid of Truth'' shown in Figure\ref{fig:pyramid}. At the apex are the expensive but essential SMEs, while the base consists of scalable but less reliable agents, creating a persistent bottleneck for creating high-quality evaluation data. Without a systematic way to manage this trade-off, several critical problems emerge:
\begin{itemize}
    \item Silent quality regressions: Without a trustworthy evaluation standard, the decision quality can silently degrade over time if not monitored.
    \item Comparison across labeling methods: We lack an objective method to compare the decision quality of different labeling vendors or to evaluate the effectiveness of different prompts for LLMs on an ``apples-to-apples'' basis. 
    \item Shifting content distributions: Production content landscapes are dynamic. New threat vectors emerge regularly and model updates alter the distribution of content, complicating the measurement of model and agent performance over time.
\end{itemize}

These challenges are central to the well-studied problem of learning from noisy labels, where statistical models infer a consensus ground truth by weighting each labeler's reliability. Such models can effectively quantify cost-quality trade-offs, for instance, by estimating how many non-expert votes are equivalent to a single expert judgment \cite{snow2008cheap}. However, our Decision Quality Evaluation Framework takes a complementary and operational approach. Rather than treating ground truth as a statistical estimate, we use experts in the highly leveraged role of curating a stable, ground truth GDS. This GDS then serves as an explicit and reliable benchmark to measure the quality of all other agents, from scalable human teams to LLMs.

This framework is especially critical for effectively managing LLMs. A key insight is that it is fundamentally easier to move the needle on LLM decision quality than on non-expert human quality. Unlike a global workforce of human agents, which is costly and slow to retrain, an LLM's decision-making process can be altered in seconds via prompt engineering. However, harnessing this agility requires a rigorous and reliable evaluation framework to measure the impact of these changes. Without one, ``prompt optimization'' remains a subjective art rather than a data-driven science.

This paper makes the following contributions:
\begin{itemize}
    \item We introduce a comprehensive framework for decision quality evaluation, featuring an automated intelligent sampling pipeline that uses propensity scores to efficiently curate and expand dataset coverage. 
    \item We define the GDS, a novel dataset construct that serves as a foundational benchmark for balancing the trade-offs between trustworthiness, coverage, and cost.
    \item We present a set of metrics to evaluate both the quality of the datasets themselves (semantic coverage and distributional divergence) and the quality of the decisions made by various labeling agents.
    \item We demonstrate the practical application of this framework at Pinterest for use cases such as prompt optimization and prevalence validation.
\end{itemize}

Through this framework, we can continuously validate the trustworthiness of enforcement decisions, set explicit quality standards for different use cases, and ensure the reproducible evaluation of all agents within our content safety ecosystem.

The remainder of the paper is structured as follows: Section 2 details our proposed framework, including the core concepts of the GDS, the metrics used to evaluate decision and dataset quality, and the overall system design. Section 3 presents several practical applications of the framework at Pinterest, demonstrating its utility in areas such as LLM prompt optimization and validating prevalence measurements. Finally, Section 4 offers concluding remarks and discusses the broader implications of our work.

\section{Framework}

\subsection{Overview}

The proposed evaluation framework is built upon the foundational concepts of trustworthiness and representativeness to enable reliable offline evaluation of decision quality.

Trustworthiness ensures that each label within our evaluation datasets accurately reflects the intent of the written policy. This eliminates ambiguity and provides a stable, ground truth reference for measuring correctness.

Representativeness is composed of two key concepts:

\begin{itemize}
    \item \textbf{Coverage}: The dataset must encompass the full spectrum of content on the platform, including diverse examples of content and critical edge cases. This ensures that performance is measured across all relevant scenarios.
    \item \textbf{Density}: This measures how closely the dataset's statistical distribution of content types aligns with the production environment.
\end{itemize}

These principles guide the practical design of our evaluation datasets, which are defined by a trade-off between four key dimensions:
\begin{itemize}
    \item Cost: The resources required to curate, label, and maintain the dataset.
    \item Size: The number of labeled examples in the dataset.
    \item Trustworthiness: The degree to which labels reliably reflect the written policy intent.
    \item Coverage: The extent to which the dataset represents the diversity of policy content and platform scenarios.
\end{itemize}

As illustrated in Figure~\ref{fig:pyramid}, different labeling sources offer varying levels of trustworthiness and scalability. SMEs produce the highest-quality labels but are costly and limited in capacity. In contrast, agents like large-scale human teams, LLMs, and Machine Learning (ML) models offer scalability at a lower cost but with reduced trustworthiness. To navigate the trade-offs defined by this ``Pyramid of Truth'' shown in Figure~\ref{fig:pyramid}, our framework's strategy is to create and maintain a foundational and high-trust dataset, the GDS.

The GDS consists of high-quality labels created and adjudicated by SMEs. Its primary purpose is to serve as the ground truth reference for label correctness. This dataset is optimized for maximum trustworthiness and broad coverage. This focus on quality and comprehensive coverage, however, results in a high cost per label, which naturally limits its size.

\begin{figure}
    \centering
    \includegraphics[width=1\linewidth]{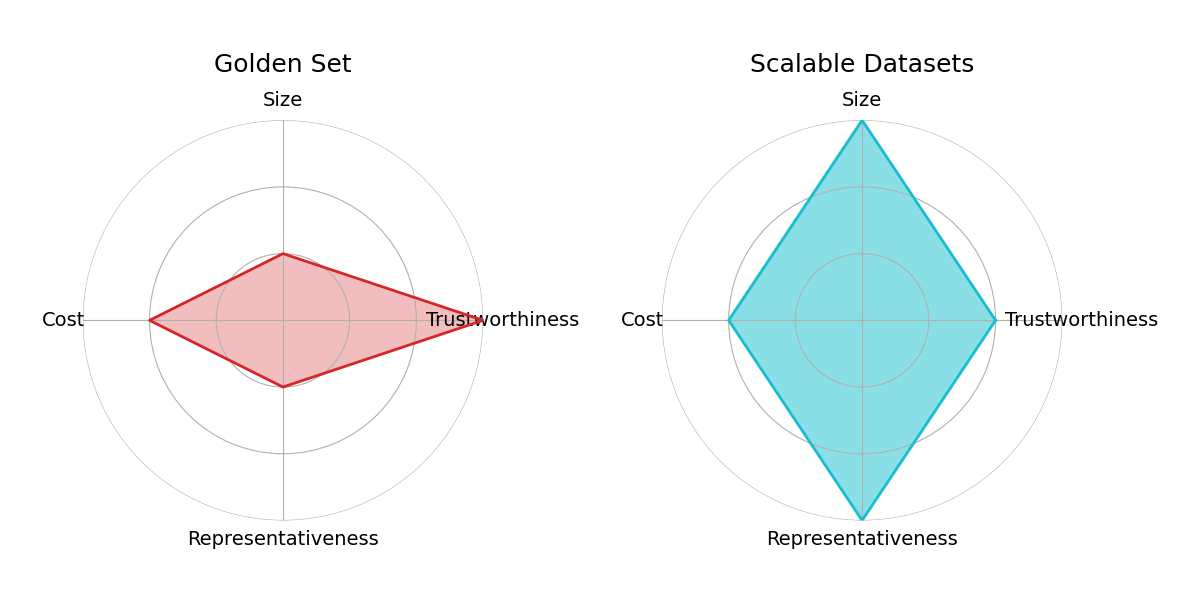}
    \caption{An illustration of design trade-offs. The GDS (left) is optimized for Trustworthiness and Coverage. In contrast, typical Scalable Datasets (right), such as those from production-scale agents, are optimized for Size and low Cost. This is a conceptual diagram as we have not actually quantified values and scales for Size, Cost, Trustworthiness and Representativeness.}
    \label{fig:gds_gcs}
\end{figure}

As Figure~\ref{fig:gds_gcs} illustrates, the design of the GDS represents an explicit choice. It prioritizes maximum trustworthiness and broad coverage over size and cost. This makes it an expensive but essential asset for reliably evaluating agent performance on complex edge cases and ensuring our interpretation of policy is measured correctly. This paper focuses exclusively on the methods for creating, maintaining, and utilizing the GDS as the foundational benchmark for all decision quality evaluation.

\subsection{Decision Quality Metrics} 
\label{sec:decision_quality_metrics} 

To evaluate the decision quality of different labelers, our framework breaks down the assessment into two components: reliability and correctness. Analyzing them together provides a powerful diagnostic for our policy feedback loop.

\textbf{Reliability} measures the consistency of judgments among different labelers. We use Cohen's Kappa $\kappa$ \cite{mchugh2012interrater}  to quantify inter-rater reliability. This metric corrects for the probability that labelers may agree by chance and provides a standardized measure of agreement. A high Kappa score indicates that our labeling guidelines are being applied consistently. As such, reliability serves as a direct proxy for the effectiveness of our internal policy guidelines and training. A low Kappa score signals that the policy is ambiguous or training has been ineffective, triggering a review.

The formula for Cohen's Kappa is
\begin{align}
    \kappa = \frac{p_o - p_e}{1 - p_e},
\end{align}
where $p_o$ is the relative observed agreement among labelers, and $p_e$ is the hypothetical probability of chance agreement.

\textbf{Correctness} is evaluated against the GDS as the ground truth. We employ a comprehensive suite of metrics to provide a holistic view of performance, as different applications may prioritize different aspects of quality. These metrics include: accuracy, precision, recall, negative precision, negative recall, False Positive Rate (FPR), False Negative Rate (FNR), informedness, and markedness.

By analyzing reliability and correctness together, we can diagnose distinct failure modes. High reliability paired with low correctness indicates that labelers are all making the same mistake consistently, which can be a systematic misunderstanding of the policy. This would require a different intervention (e.g., broad retraining on a specific subpolicy or concept) than low reliability. 

\subsection{Dataset Metrics} 
\label{sec:dataset_metrics} 

To ensure our GDS is a robust foundation for evaluation, we must also measure its intrinsic quality, specifically its representativeness. The GDS must be comprehensive enough to ensure that evaluations against it are meaningful. We measure this through two key metrics: ``semantic coverage'' and ``distributional divergence.''

\textbf{Semantic Coverage} is used to measure the diversity of our dataset. We define this metric using semantic IDs generated from our internal PinCLIP image embeddings \cite{beal2026pinclip}, which build upon foundational work in multi-modal models, such as CLIP \cite{radford2021learning} and SigLIP \cite{zhai2023sigmoid}. These embeddings are quantized using a Residual-Quantized Variational Autoencoder (RQ-VAE) \cite{lee2022autoregressive, singh2024better}, which produces a hierarchical sequence of discrete codes. The first layer of this VAE has a codebook of $2^8=256$ unique codes, each representing a high-level semantic cluster.

Let the set of all 256 possible first-layer codes be $C = \{c_1, c_2, \dots, c_{K}\}$ and $K=256$, where each code $c_j$ represents a high-level semantic cluster. Given a dataset $D$ and a function $f: D \to C$ that maps each item in the dataset to its first-layer code, we define the set of observed codes as $C_{obs} = \{ f(d) \mid d \in D \}$. The coverage of the dataset is then formally defined as the percentage of these foundational semantic codes that are observed:
\begin{align}
    \text{Semantic Coverage} = \frac{|C_{obs}|}{|C|}  = \frac{|C_{obs}|}{256}
\end{align}

A higher coverage value signifies that the dataset encompasses a broader range of the visual concepts captured by the embedding space, which in turn guides our data augmentation strategy.

\textbf{Distributional Divergence} quantifies how the statistical profile of our dataset compares to the data distribution in our live production environment. While high coverage is a goal for the GDS, it is often intentionally not distributionally representative of production, as we oversample for rare content and edge cases. This metric allows us to understand and quantify that difference.

We define two probability distributions over the $K=256$ first-layer semantic clusters, $C = \{c_1, \dots, c_K\}$:

\begin{itemize}
    \item The Production Distribution ($P_{prod}$): This is the empirical probability distribution of semantic clusters calculated from a large representative sample of our live production traffic. For each cluster $c_j$, its probability $P_{prod}(c_j)$ is its relative frequency in the production environment.
    \item  The Dataset Distribution ($P_{D}$): This is the empirical probability distribution of semantic clusters in our dataset $D$. For each cluster $c_j$, its probability $P_D(c_j)$ is its relative frequency in the dataset.
\end{itemize}

We quantify the divergence between our dataset distribution ($P_D$) and the production distribution ($P_{prod}$) using the Jensen-Shannon Divergence (JSD) \cite{lin2002divergence}. JSD is a symmetric and bounded metric that measures the statistical distance between probability distributions. It is defined as follows:

\begin{align}
    JSD(P_D || P_{prod}) = \frac{1}{2} D_{KL}(P_D || M) + \frac{1}{2} D_{KL}(P_{prod} || M),
\end{align}
where $M = \frac{1}{2}(P_D + P_{prod})$ is the pointwise mean of the two distributions, and $D_{KL}(P || Q) = \sum_{j=1}^{K} P(c_j) \log_2 \frac{P(c_j)}{Q(c_j)}$ denotes the Kullback-Leibler divergence.

The JSD score, calculated with a base-2 logarithm, is bounded between 0 (for identical distributions) and 1 (for maximally different distributions).

Together, coverage and distributional similarity provide a comprehensive profile of our dataset's content, quantifying both its conceptual breadth and its statistical relationship to the production environment.

\subsection{Design} 

\begin{figure}[ht]
    \centering
    \includegraphics[width=\linewidth]{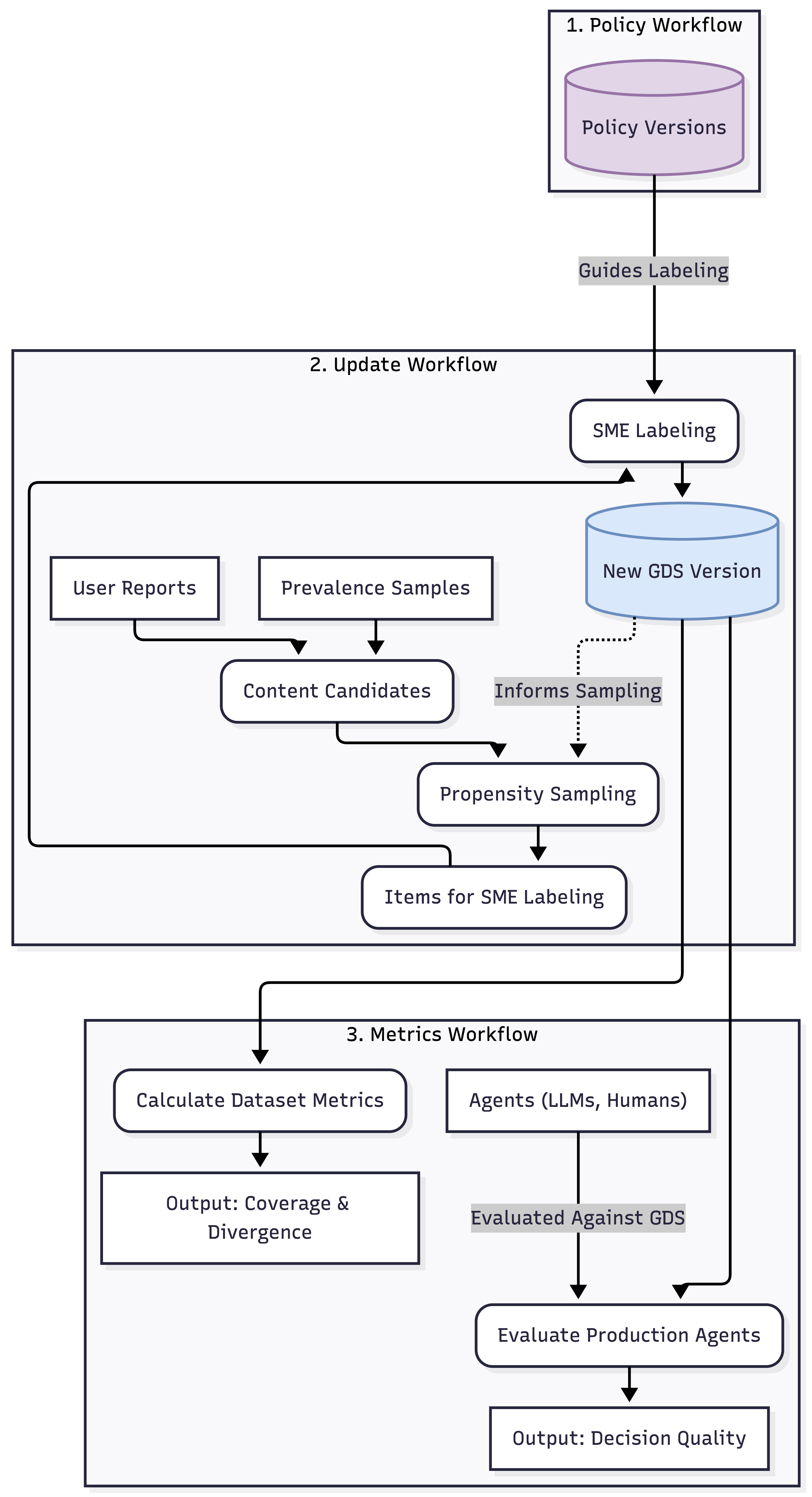}
    \caption{The automated evaluation framework, showing the three core workflows. The Update Workflow creates a new GDS version, which triggers the Metrics Workflow. The resulting coverage metric is fed back to inform future sampling, creating an intelligent loop.}
    \label{fig:workflow}
\end{figure}

The framework is operationalized as an automated system built around three modular and interconnected workflows that run on a regular cadence shown in Figure~\ref{fig:workflow}. This cyclical process ensures our evaluation benchmarks and quality metrics remain current with the dynamic nature of our platform.

 \begin{itemize} 
 \item \textbf{Policy workflow}: This workflow captures the policy taxonomy and its versions. It provides a canonical lookup for every policy, its associated labels and the entity type to which it applies. This ensures that all labels and decisions are tied to a specific, immutable version of the policy. As the diagram shows, its output directly guides labeling to ensure that all decisions made are consistent with the current and correct policy interpretation. 
 
\item \textbf{Update workflow}: This workflow is responsible for creating a New GDS Version. It ingests content candidates, intelligently selects a subset for labeling, and publishes the result as an immutable and versioned dataset.
\begin{enumerate}
    \item Ingests Candidates: It sources content from various streams, such as User Reports and Prevalence Samples, to create a pool of Content Candidates for potential inclusion in the GDS.
    \item Selects for Labeling using Propensity Sampling: To efficiently increase the coverage of the GDS, we employ a model-driven active sampling strategy. As the diagram illustrates, this Propensity Sampling step is guided by two inputs: the pool of Content Candidates and, the existing GDS Version, via the ``Informs Sampling'' feedback loop. Specifically, an XGBoost model is trained to predict the propensity score $p(\text{item} \in \text{GDS} \mid \text{embedding})$ based on an item's PinCLIP embeddings. For the next labeling batch, we use inverse propensity sampling, prioritizing candidates with a low score. This actively seeks out novel and underrepresented items, ensuring our limited SME Labeling budget is spent on content that most effectively expands the GDS's coverage.
    \item Publishes New Dataset Version: The selected Items for SME Labeling are sent to SMEs. Once their labels are collected, the items are integrated into the GDS to publish a New GDS Version.

\end{enumerate}
\item \textbf{Metrics workflow}: This automated workflow runs immediately after a New GDS Version is published, closing the evaluation loop. It takes the new GDS as input and computes two categories of metrics in parallel: 
\begin{enumerate} 
\item Dataset Metrics: The workflow first performs the Calculate Dataset Metrics step. This measures the intrinsic quality of the New GDS Version, producing the final Output: Coverage \& Divergence (as described in Section \ref{sec:dataset_metrics}). 
\item Decision Quality Metrics: Simultaneously, production Agents (LLMs, Humans) are evaluated. As the diagram shows, they are Evaluated Against GDS in the Evaluate Production Agents step. This process generates the final Output: Decision Quality metrics (e.g., accuracy, recall, as described in Section \ref{sec:decision_quality_metrics}). 
\end{enumerate}

\end{itemize}

This versioned and modular design ensures that our evaluations are reproducible and can adapt to the evolving landscape of content safety policies and enforcement mechanisms.

\section{Applications}

The principles of the proposed framework have been applied at Pinterest to measure decision quality in key content safety areas, enabling a data-driven approach to policy enforcement. This section details three key applications within adult dontent as an example. To evaluate our labelers, we curated an adult content GDS. The GDS is well-balanced between positive and negative examples and has wide semantic coverage of the problem space, ensuring it provides a robust and comprehensive benchmark.

\subsection{Benchmarking Agent Quality and Driving LLM Optimization}

\begin{table*}[ht]
\centering
\caption{
    Relative decision quality metrics for different labelers measured on an adult content GDS.
    All values are shown as \textbf{percentage point differences} relative to the ``1x human'' baseline.
    For performance metrics (e.g., Acc., F1), positive values indicate an improvement over the baseline.
    For error metrics (FPR, FNR), negative values indicate an improvement (a reduction in errors).
}
\label{tab:relative_decision_quality}
\resizebox{\textwidth}{!}{%
\begin{tabular}{@{}lcccccccccccc@{}}
\toprule
\textbf{Labeler} & \textbf{Acc.} & \textbf{Prec.} & \textbf{Recall} & \textbf{F1} & \textbf{Neg. Prec.} & \textbf{Neg. Rec.} & \textbf{FPR} & \textbf{FNR} & \textbf{Inform.} & \textbf{Marked.} & \textbf{Pred. Pos. Frac.} & \textbf{Pos. Prev.} \\
\midrule
\textbf{1x human (Baseline)} & \textbf{0.0} & \textbf{0.0} & \textbf{0.0} & \textbf{0.0} & \textbf{0.0} & \textbf{0.0} & \textbf{0.0} & \textbf{0.0} & \textbf{0.0} & \textbf{0.0} & \textbf{0.0} & \textbf{0.0} \\
3x human majority & +3.6 & +2.5 & +4.0 & +3.4 & +4.1 & +2.8 & -2.8 & -4.0 & +6.8 & +6.7 & +1.0 & -0.2 \\
\midrule
Gemini 2.5 flash (min FPR) & -6.7 & -18.1 & +17.3 & -1.2 & +10.0 & -39.9 & +39.9 & -17.3 & -22.6 & -7.9 & +26.9 & +0.3 \\
Gemini 2.5 flash (balanced) & -10.9 & -23.0 & +22.6 & -2.7 & +18.7 & -57.2 & +57.2 & -22.6 & -34.6 & -4.2 & +37.3 & +0.3 \\
Gemini 2.5 pro (min FPR) & -7.0 & -19.0 & +19.5 & -1.0 & +13.5 & -43.5 & +43.5 & -19.5 & -24.1 & -5.3 & +29.7 & +0.3 \\
Gemini 2.5 pro (balanced) & -6.9 & -19.8 & +22.5 & -0.5 & +20.5 & -47.7 & +47.7 & -22.5 & -25.3 & +0.8 & +33.2 & +0.3 \\
GPT-4.1 (balanced) & -3.2 & -10.9 & +9.0 & -0.5 & +3.4 & -20.0 & +20.0 & -9.0 & -11.0 & -7.4 & +13.8 & +0.3 \\
GPT-4o (balanced) & -4.1 & -10.8 & +6.6 & -1.5 & +1.0 & -18.9 & +18.9 & -6.6 & -12.3 & -9.7 & +11.9 & +0.3 \\
GPT-5 (min FPR) & -5.8 & -1.0 & -10.3 & -6.8 & -7.8 & +0.5 & -0.5 & +10.3 & -9.8 & -8.7 & -6.0 & +0.4 \\
GPT-5 (balanced) & +0.9 & -8.2 & +13.5 & +3.1 & +9.9 & -16.5 & +16.5 & -13.5 & -3.0 & +1.8 & +15.0 & +0.3 \\
\bottomrule
\end{tabular}%
}
\end{table*}

A primary challenge in large-scale content moderation is the objective comparison of diverse labeling agents. This process of empirically evaluating and comparing the performance of different human-AI teams is a cornerstone of the research field dedicated to the science of human-AI decision making \cite{lai2021towards}. To address this, we utilized the adult content GDS to establish the unified quality benchmark across both human teams and LLM-based systems. This provides a quantitative basis for agent selection, performance tracking, and prompt engineering. The results of this analysis are presented in Table~\ref{tab:relative_decision_quality}, with all numbers shown as percentage differences relative to the 1x non-expert human performance baseline. Performance comparisons reflect research benchmarks used for internal optimization and production systems employ ensemble approaches not represented by individual agent metrics. A key takeaway from Table~\ref{tab:relative_decision_quality} is that the LLMs demonstrate quality on par with a single non-expert human, but a gap still remains to reach SME-level quality.

Beyond static comparisons, a key application is using the framework to drive the optimization of LLM behavior through prompt engineering. In this workflow, the GDS serves as the fixed evaluation set for iterative development. Engineers can modify a prompt, test it against the GDS, and receive immediate, trustworthy feedback. This transforms prompt development from a subjective art into a quantitative science, enabling teams to establish formal exit criteria. For example, a team could define a quality bar relative to a baseline, such as achieving an Informedness score of 5\% higher than a non-expert human, which signals a prompt is ready for production. The min FPR prompt and the balanced prompt in Table~\ref{tab:relative_decision_quality} are the outputs of two such optimization cycles, each targeting a different exit criterion. The framework is efficient in guiding development toward a specific and measurable target.

The benchmarking results reveal critical performance trade-offs between different agent types. For instance, the 3x human majority configuration exemplifies a high-certainty agent. It improves upon the single human baseline across key metrics, achieving a 2.5\% higher precision and a 4.0\% higher recall, while simultaneously reducing the FPR by 2.8\%. This indicates it is highly reliable when identifying content as in-scope policy and rarely makes mistakes on out-of-scope content, making it a trustworthy baseline. In contrast, the Gemini 2.5 pro with a balanced prompt demonstrates a clear trade-off: it delivers a substantial 22.5\% gain in recall over the baseline, but this comes at the cost of a much higher FPR, which increases by 47.7\%. Our framework quantifies this trade-off precisely, allowing for a principled labeler selection based on task-specific error costs.

This rigorous evaluation extends to the crucial task of cost-performance optimization when selecting a model. Different models, even within the same family, come with distinct latency and cost profiles. Our framework provides the quantitative data needed to make informed financial decisions. For instance, when comparing GPT-4.1 and GPT-4o, our results indicate that GPT-4.1 offers a superior performance profile for our needs. Compared to the human baseline, it achieves a higher gain in recall (+9.0\% vs. +6.6\%) and, critically, a smaller performance deficit in informedness (-11.0\% vs. -12.3\%). This quantified performance gap is the critical input required to determine if the marginal gain from one model justifies its associated cost. By translating abstract model capabilities into concrete metrics, our framework moves the decision from a subjective choice to a quantitative cost-benefit analysis, allowing us to confidently select the model that provides the optimal trade-off for any given application and budget.

\subsection{Managing Policy Update}

The dynamic nature of Content-Safety requires that content policies constantly evolve to address emerging changes and the latest trends. A key challenge is not just adapting to these changes, but first quantitatively understanding the change itself. Our framework provides a novel methodology for this, centered on dual-labeling the GDS to characterize the ``policy delta''. 

\begin{figure}
    \centering
    \includegraphics[width=\linewidth]{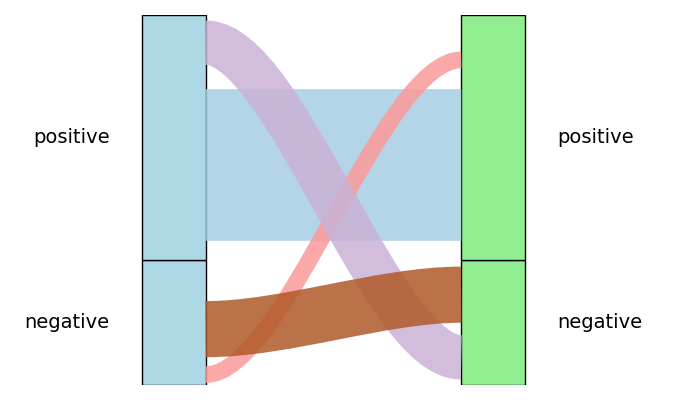}
    \caption{The Sankey diagram visualizing the policy delta on the GDS, showing the flow of items from old SME labels to new SME labels after a policy update.}
    \label{fig:sankey}
\end{figure}

Stage 1: Characterizing the Policy Delta. To understand the precise impact of a policy update, the existing GDS is relabeled by SMEs under the new policy guidelines. This creates two sets of ground truth labels for the same content: \texttt{GDS\_labels\_v1} (old policy) and \texttt{GDS\_labels\_v2} (new policy). By comparing these two label sets, we can generate a ``policy delta'' analysis, often visualized as a Sankey diagram (Figure~\ref{fig:sankey}). This diagram visualizes this policy delta, showing the flow of items between their labels under the old and new policies (e.g., from ``negative'' to ``positive'') This provides product and policy teams with a clear and quantitative map of the change's impact before any enforcement agents are modified.

Stage 2: Evaluating Agent Adaptation. Once the new policy is established, the framework is used to measure the decision quality of all enforcement agents, such as human agents and new LLM prompts, against the new standard. In this process, the agents' decisions on the GDS items are evaluated solely against the new SME labels (\texttt{GDS\_labels\_v2}). 

This provides a clear, unambiguous benchmark of each agent's compliance with the current, correct policy. This comprehensive evaluation is sufficient to detect regressions, as an agent that has degraded on still-valid rules will exhibit lower overall performance against the new label set. More broadly, this evaluation process culminates in a full re-benchmarking of the entire agent ecosystem, including 1x human, 3x human majority, and various LLM configurations, against the new policy ground truth. The result is the decision quality analysis presented in Table~\ref{tab:relative_decision_quality}, ensuring our understanding of agent performance and their associated cost-tradeoffs remains current as our safety policies evolve.

\subsection{Prevalence Validation}

Accurately estimating the prevalence of rare policy content at platform scale is a critical yet challenging task. These metrics inform strategic policy and business decisions, and their integrity is paramount. Our platform recently transitioned from using costly 3x human majority labels to a more scalable LLM-based approach for this task. This shift resulted in over 30x in cost savings and a 10x reduction in labeling turnaround time.

This transition from a trusted human consensus to an automated system necessitates a rigorous and continuous validation framework to ensure the LLM's quality does not degrade over time, which would silently bias our prevalence metrics. To achieve this, we use the GDS to implement a continuous monitoring system with two distinct tracks, each designed to detect a different type of potential failure.

First, we monitor for content drift by regularly evaluating the LLM's performance on new content added with each GDS release. This process answers the critical question: ``Is the model's quality degrading as real-world content patterns evolve?'' By focusing the evaluation on the newest GDS items, we can quickly detect if our LLM is failing to generalize to emerging adversarial trends. A significant drop in performance on this new data triggers an immediate investigation and potential prompt updates. This prevents evolving content from invalidating our measurement system.

Second, we validate system stability. This process tests for technical issues by periodically re-evaluating the same LLM and prompt against a specific GDS version that was used to establish the initial decision quality baseline for the current policy. In this task, both the data and the model configuration are constant. Therefore, any significant change in decision quality metrics would not be due to content changes, but would instead indicate a potential system-level issue, such as a pipeline bug, an unintended change in a dependency, or non-determinism in the inference stack. This allows us to isolate engineering instability from model quality degradation caused by evolving content.

These two processes, working in concert, provide a comprehensive safeguard for our prevalence estimation pipeline. By continuously validating our LLM agent against both evolving content (content drift) and a fixed benchmark (system stability), the framework ensures the integrity of our most critical platform-wide metrics over the long term.

\section{Conclusions}
In this paper, we presented a comprehensive framework for the systematic evaluation of content moderation decisions at scale. Our work is motivated by the need to manage the trade-offs between cost, trustworthiness, and scale, while taking advantage of the agility of modern LLMs. Our primary contribution is a holistic system built around a high-trust and expert-curated GDS, an intelligent sampling pipeline for dataset curation, and a set of robust, automated workflows for continuous evaluation.

Through practical applications at Pinterest, we demonstrated that the framework provides a powerful solution to several real-world challenges. We have shown its ability to:
\begin{itemize}
    \item Quantify performance trade-offs between human agents and various LLM configurations, enabling data-driven cost-benefit analysis.
    \item Provide a rigorous methodology for prompt engineering, transforming it from a subjective art into a quantitative science with clear exit criteria.
    \item Characterize and manage the entire policy evolution lifecycle, from understanding the impact of a policy change to measuring agent adaptation.
    \item Safeguard the integrity of critical prevalence metrics by continuously monitoring for both content drift and system instability.
\end{itemize}

The broader implication of this work is a blueprint for operationalizing responsible AI development in the Content-Safety domain. By creating a reliable, reproducible, and intelligent evaluation system, we provide a pathway for organizations to move from subjective quality assessments to a continuous, data-driven practice. 
A promising direction for future work is to leverage the GDS to build a new class of large-scale evaluation datasets. These datasets would intentionally sacrifice a degree of trustworthiness for massive gains in size and cost-efficiency by using the GDS as a benchmark to calibrate labels from more scalable sources. The availability of such datasets, which we term the ``Gold-Certified Set'', would enable a broader range of applications that require large-scale data, such as offline ML model evaluation.

\begin{acks}
This work cannot be accomplished without the help from Minli Zang, Tony Paek, Xiaohan Yang, Benjamin Thompson, and Monica Bhide. We would like to thank them for their support and contributions throughout this project.
\end{acks}

\bibliographystyle{ACM-Reference-Format}
\bibliography{sample-base}

\appendix

\end{document}